%% file: main.tex
\documentclass[sigconf]{acmart}

\def\BibTeX{{\rm B\kern-.05em{\sc i\kern-.025em b}\kern-.08emT\kern-.1667em\lower.7ex\hbox{E}\kern-.125emX}}
    
\copyrightyear{2019}
\acmYear{2019}
\setcopyright{acmlicensed}
\acmConference[The Web Conference '19]{International Workshop on Misinformation, Computational Fact-Checking and Credible Web}{May 14, 2019}{San Francisco, CA}
\acmBooktitle{International Workshop on Misinformation, Computational Fact-Checking and Credible Web, May 14, 2019, San Francisco, CA}
\acmPrice{15.00}
\acmDOI{10.1145/1122445.1122456}
\acmISBN{978-1-4503-9999-9/18/06}

\usepackage{times}  
\usepackage{helvet}  
\usepackage{courier}  
\usepackage{url}  
\usepackage{graphicx}  
\usepackage{color-edits}
\usepackage{amsmath}
\usepackage{balance}
\usepackage{subcaption}
\usepackage{mwe}
\usepackage{booktabs}
\usepackage{multirow}
\usepackage{tikz}

\addauthor{ab}{blue}  
\addauthor{ad}{red}
\addauthor{ef}{green}
\addauthor{ll}{cyan}

\begin{document}

\title[Red Bots Do It Better]{Red Bots Do It Better:\\Comparative Analysis of Social Bot Partisan Behavior}

\author[L. Luceri]{Luca Luceri}
\thanks{L. Luceri \& A. Deb contributed equally to this work.}
\additionalaffiliation{%
  \institution{USC Information Sciences Institute}
  \city{Marina del Rey}
  \state{CA}
}
\affiliation{%
  \institution{University of Applied Sciences and Arts of Southern Switzerland, and University of Bern}
  \city{Manno}
  \state{Switzerland}
  \postcode{6928}
}
\email{luca.luceri@supsi.ch}

\author[A. Deb]{Ashok Deb}
\affiliation{%
  \institution{USC Information Sciences Institute}
  \city{Marina del Rey}
  \state{CA}
  \postcode{90292}
}
\email{ashok@isi.edu}

\author[A. Badawy]{Adam Badawy}
\affiliation{%
  \institution{USC Information Sciences Institute}
  \city{Marina del Rey}
  \state{CA}
  \postcode{90292}
}
\email{badawy@isi.edu}

\author[E. Ferrara]{Emilio Ferrara}
\affiliation{%
  \institution{USC Information Sciences Institute}
  \city{Marina del Rey}
  \state{CA}
  \postcode{90292}
}
\email{emiliofe@usc.edu}


\begin{abstract}
Recent research brought awareness of the issue of bots on social media and the significant risks of mass manipulation of public opinion in the context of political discussion.
In this work, we leverage Twitter to study the discourse during the 2018 US midterm elections and analyze social bot activity and interactions with humans. We collected 2.6 million tweets for 42 days around the election day from nearly 1 million users. 
We use the collected tweets to answer three research questions: $(i)$ \textit{Do social bots lean and behave according to a political ideology?} $(ii)$ \textit{Can we observe different strategies among liberal and conservative bots?} $(iii)$ \textit{How effective are bot strategies?} 

We show that social bots can be accurately classified according to their political leaning and behave accordingly. Conservative bots share most of the topics of discussion with their human counterparts, while liberal bots show less overlap and a more inflammatory attitude. We studied bot interactions with humans and observed different strategies.
Finally, we measured bots embeddedness in the social network and the effectiveness of their activities. 
Results show that conservative bots are more deeply embedded in the social network and more effective than liberal bots at exerting influence on humans.


\end{abstract}

\maketitle

\input{src/introduction.tex}
\input{src/data.tex}

\input{src/methodology.tex}

\input{src/results.tex}

\input{src/discussion.tex}
\input{src/conclusion.tex}

\footnotesize{\bigskip
\textbf{Acknowledgements}. 
The authors gratefully acknowledge support by the Air Force Office of Scientific Research (award \#FA9550-17-1-0327). 
L. Luceri is funded by the Swiss National Science Foundation (SNSF) via the CHIST-ERA project \textit{UPRISE-IoT}.
}
\balance
\bibliographystyle{ACM-Reference-Format}
\bibliography{ref}

\end{document}

%% file: src/introduction.tex
\section{Introduction}
During the last decade, social media have become the conventional communication channel to socialize, share opinions, and access the news.
Accuracy,  truthfulness, and  authenticity of the shared content are necessary ingredients to maintain a healthy online discussion.
However, in recent times, social media have been dealing with a considerable growth of false content and fake accounts.
The resulting wave of misinformation (and disinformation) highlights the pitfalls of social media networks and their potential harms to several constituents of our society, ranging from politics to public health.

In fact, social media networks have been used for malicious purposes to a great extent \cite{ferrara2015manipulation}. 
Various studies raised awareness about the risk of mass manipulation of public opinion, especially in the context of political discussion.
Disinformation campaigns~\cite{persily20172016, howard2017junk, shu2017fake, ferrara2017disinformation, vosoughi2018spread, badawy2018falls, guess2019less, bovet2019influence, scheufele2019science, Grinberg2019} and social  bots~\cite{bessi2016social, woolley2017computational, varol2017online,  monsted2017evidence, pozzana2018measuring, boichak2018automated, shao2018spread, yang2019arming} have been indicated as factors contributing to social media manipulation.

The 2016 US Presidential election represents a prime example of the significant perils of mass manipulation of political discourse. 
\citet{Badawy2018} studied the Russian interference in the election and the activity of Russian trolls on Twitter.  \citet{im2019still} suggested that troll accounts are still active to these days.
The presence of social bots does not show any sign of decline \cite{yang2019arming,deb2019bots} despite the attempts from social network providers to suspend  suspected, malicious accounts.
Various research efforts have been focusing on the analysis, detection, and countermeasures development against social bots. \citet{ferrara2016rise} highlighted the consequences associated with bot activity in social media. The online conversation related to the 2016 US presidential election was further examined  \cite{bessi2016social} to quantify the extent of social bots activity. More recently,
\citet{stella2018bots} discussed bots' strategy of targeting influential humans to manipulate online conversation during the Catalan referendum for independence, whereas \citet{shao2018spread} analyzed the role of social bots in spreading articles from low credibility sources.
\citet{deb2019bots} focused on the 2018 US Midterms elections with the objective to find instances of voter suppression. 

In this work, we investigate social bots behavior by analyzing their activity, strategy, and interactions with humans. 
We aim to answer the following research questions (RQs) regarding social bots behavior during the 2018 US Midterms election. 

\begin{itemize}
    \item[\textbf{RQ1}:]  \textit{Do social bots lean and behave according to a political ideology?} We investigate whether social bots can be classified based on their political inclination into liberal or conservative leaning. Further, we explore to what extent they act similarly to the corresponding human counterparts.

    \item[\textbf{RQ2}:]  \textit{Can we observe different strategies among liberal and conservative bots?} We examine the differences between social bot strategies to mimic humans and infiltrate political discussion. For this purpose, we measure bot activity in terms of volume and frequency of posts, interactions with humans, and embeddedness in the social network.
    \item[\textbf{RQ3}:]  \textit{Are bot strategies effective?} We introduce four metrics to estimate the effectiveness of bot strategies and to evaluate the degree of human interplay with social bots.

\end{itemize}

We leverage Twitter to capture the political discourse during the 2018 US midterm elections. We collected 2.6 million tweets for 42 days around  election day from nearly 1 million users. 
We then explore  collected  data and  attain the following findings:
\begin{itemize}
    \item We show that social bots are embedded in each political side and behave accordingly. Conservative bots abide by the topic discussed by the human counterpart more than liberal bots, which in turn exhibit a more provocative attitude. 
    \item We examined bots' interactions with humans and observed different strategies. Conservative bots stand in a more central   social network position, and divide their interactions between humans and other conservative bots, whereas liberal bots focused mainly on the interplay with the human counterparts.
    \item We measured the effectiveness of these strategies and recognized the strategy of  conservative bots as the most effective in terms of influence exerted on human users.
\end{itemize}

%% file: src/data.tex
\section{Data}

In this study, we use Twitter to investigate the partisan behavior of malicious accounts during the 2018 US midterm elections.
For this purpose, we carried out a data collection from the month prior (October 6, 2018) to two weeks after (November 19, 2018) the day of the election.
We kept the collection running after the election day as several races remained unresolved.
We employed the Python module \textit{Twyton} to collect tweets through the Twitter Streaming API using the following keywords as a filter: \textit{2018midtermelections}, \textit{2018midterms}, \textit{elections}, \textit{midterm}, and \textit{midtermelections}.
As a result, we gathered 2.7 million tweets, whose IDs are publicly available for download.\footnote{\url{https://github.com/A-Deb/midterms}}
From this set, we first removed any duplicate tweet, which may have been captured by accidental redundant queries to the Twitter API.
Then, we excluded all the tweets not written in English language.
Despite the majority of the tweets were in English, and to a lesser degree in Spanish (3,177 tweets), we identified 59 languages in the collected data. 
Thus, we inspected tweets from other countries and removed them as they were out of the context of this study.
In particular, we filtered out tweets related to the Cameroon election, the Democratic Republic of the Congo election, the Biafra call for Independence, democracy in Kenya (\#democracyKE), to the two major political parties in India (BJP and UPA), and college midterm exams.
Overall, we retain nearly 2.6 millions tweets, whose aggregate statistics are reported in Table \ref{tab:datasets_stats}. 

\begin{table}[t]
\centering \small
\caption{Dataset statistics}
\label{tab:datasets_stats}
\begin{tabular}{|l|c|c|c|}
\hline
    Statistic & Count   \\ 
    
\hline
\# of Tweets	&		452,288	\\
\# of Retweets	&		1,869,313	\\
\# of Replies	&		267,973	\\
\# of Users	&	997,406	\\
\hline
\end{tabular} 
\vspace{-.5cm}
\end{table}

%% file: src/methodology.tex
\section{Methodology}

\subsection{Bot Detection}
Nowadays, bot detection is a fundamental asset for understanding social media manipulation and, more specifically, to reveal malicious accounts. 
In the last few years, the problem of detecting automated accounts gathered both attention and concern  \cite{ferrara2016rise}, also bringing a wide variety of approaches to the table \cite{subrahmanian2016darpa,kudugunta2018deep,chavoshi2016debot,chen2018unsupervised}. 
While increasingly sophisticated techniques keep emerging \cite{kudugunta2018deep}, in this study, we employ the widely used \textit{Botometer}.\footnote{\url{https://botometer.iuni.iu.edu/}}

Botometer is a machine learning-based tool developed by Indiana University \cite{davis2016botornot,varol2017online} to detect social bots in Twitter.
It is based on an ensemble classifier \cite{breiman2001random} that aims to provide an indicator, namely \textit{bot score}, used to classify an account either as a bot or as a human.
To feed the classifier, the Botometer API extracts about 1,200 features related to the Twitter account under analysis.
These features fall in six broad categories and characterize the account's profile, friends, social network, temporal activity patterns, language, and sentiment.
Botometer outputs a bot score: the lower the score, the higher the probability that the user is human. 
In this study we use version \textit{v3} of Botometer, which brings some innovations, as detailed in \cite{yang2019arming}. Most importantly, the bot scores are now rescaled (and not centered around 0.5 anymore) through a non-linear re-calibration of the model.

\begin{figure}[t] 
\centering
  \includegraphics[width=.9\columnwidth]{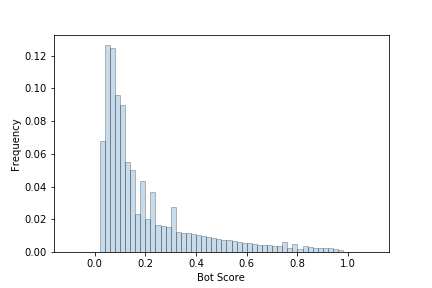}
  \caption{Bot score distribution }
  \label{fig:bot_score_dist}
\vspace{-.5cm}
\end{figure}

In Figure \ref{fig:bot_score_dist}, we depict the bot score distribution of the 997,406 distinct users in our datasets. 
The distribution exhibits a right skew: most of the probability mass is in the range [0, 0.2] and some peaks can be noticed around 0.3. 
Prior studies used the 0.5 threshold to separate humans from bots. However, according to the re-calibration introduced in Botometer \textit{v3} \cite{yang2019arming}, along with the emergence of increasingly more sophisticated bots, we here lower the bot score threshold to 0.3 (i.e., a user is labeled as a bot if the  score is above 0.3). This threshold corresponds to the same level of sensitivity setting of 0.5 in prior versions of Botometer (cf. Fig 5 from \cite{yang2019arming}). 

According to this choice, we classified 21.1\% of the accounts as bots, which in turn generated 30.6\% of the tweets in our data set.
Overall, Botometer did not return a score for 35,029 users that corresponds to 3.5\% of the accounts.
We used the Twitter API to further inspect them. Interestingly, 99.4\% of these accounts were suspended by Twitter, whereas the remaining percentage of users protected their tweets  turning on the privacy settings of their accounts.

\subsection{Political Ideology Inference}
In parallel to the bot detection analysis, we examine the political leaning of both bots and humans in our dataset.
To classify users based on their political ideology, we rely on the political leaning of the media outlets they share. We make use of a list of partisan media outlets released by third-party organizations, such as AllSides\footnote{\url{https://www.allsides.com/media-bias/media-bias-ratings}} and Media Bias/Fact Check.\footnote{\url{https://mediabiasfactcheck.com/}}
We combine liberal and liberal-center media outlets into one list (composed of 641 outlets) and conservative and conservative-center into another (composed of 398 outlets). 
To cross reference these media URLs with the URLs in the Twitter dataset, we need to get the expanded URLs for most of the links in the dataset, as most of them are shortened. 
However, this process is quite time-consuming, thus, we decided to rank the top 5,000 URLs by popularity and retrieve the long version only for those. These top 5,000 URLs accounts for more than 254K, or more than 1/3 of all the URLs in the dataset. 
After cross-referencing the 5,000 extended URLs with the media URLs, we observe that 32,115  tweets in the dataset contain a URL that points to one of the liberal media outlets and 25,273 tweets with a URL pointing to one of the conservative media outlets. 

To label Twitter accounts as liberal or conservative, we use a polarity rule based on the number of tweets they produce with links to liberal or conservative sources. Thereby, if an account has more tweets with URLs pointing to liberal sources, it is labeled as liberal and vice versa. Although the overwhelming majority of accounts include URLs that are either liberal or conservative, we remove any account that has equal number of tweets from each side. Our final set of labeled accounts includes 38,920 users.

Finally, we use label propagation to classify the remaining accounts in a similar way to previous work (\textit{cf.} \cite{Badawy2018}). 
For this purpose, we construct a social network based on the retweets exchanged between users.
The nodes of the retweet network are the users, which are connected by a direct link if one user retweeted a post of another user.  
To validate results of the label propagation algorithm, we apply a stratified cross (5-fold) validation to a set composed of 38,920 seed accounts.
We train the algorithm using 80\% of the seeds and we evaluate the performance on the remaining 20\%.
Finally, we compute precision and recall by reiterating the validation of the 5-folds. Both precision and recall scores show value around 0.89 and validate the proposed approach.
Moreover, since we combine liberal and liberal-center into one list (same for conservatives), we can see that the algorithm is not only labeling the far liberal or conservative correctly, which is a relatively easier task, but it is performing well on the liberal/conservative center as well.

\subsection{Bot Activity Effectiveness}
We next introduce four metrics to estimate bot effectiveness and, at the same time, measure to what extent humans rely upon, and interact with the content generated by social bots.
Thereby, we propose the following metrics:
\begin{itemize}
    \item \textit{Retweet Pervasiveness} ($RTP$) measures the intrusiveness of bot-generated content in human-generated retweets:
    \begin{equation}
 RTP=\frac{\mbox{no. of human retweets from bot tweets}}{\mbox{no. of human retweets}}
  \end{equation}
    \item \textit{Reply Rate} ($RR$) measures the percentage of replies given by humans to social bots:
    \begin{equation}
 RR=\frac{\mbox{no. of human replies to bot tweets}}{\mbox{no. of human replies}}
  \end{equation}
    \item \textit{Human to Bot Rate} ($H2BR$) quantifies human interaction with bots over all the human activities in the social network:
     \begin{equation}
H2BR=\frac{\mbox{no. of humans interaction with bots}}{\mbox{no. of humans activity}}\mbox{,}
  \end{equation}
where the numerator counts for human replies/retweets to/of bots generated content, while the denominator is the sum of the number of human tweets, retweets, and replies.     
    \item \textit{Tweet Success Rate} ($TSR$) is the percentage of tweets generated by bots that obtained at least one retweet by a human:
 \begin{equation}
TSR=\frac{\mbox{no. of tweet retweeted at least once by a human}}{\mbox{no. of bots tweets}}
  \end{equation}
\end{itemize}

%% file: src/results.tex
\section{Results}
Next, we address the research questions discussed in the Introduction.
We examine social bot partisanship and, accordingly, we analyze bots' strategies and measure the effectiveness of their actions.

\subsection{RQ1: Bot Political Leaning}
\begin{table}[t]
\caption{Users and tweets statistics}
\label{tab: stats}
\centering\small
\begin{subtable}{\columnwidth}
\centering
\begin{tabular}{ccc}
\hline

& Liberal    & Conservative \\
\hline
Humans      &386,391 (38.7\%)   & 122,761 (12.3\%)     \\
Bots      &82,118 (8.2\%)     & 49,488 (4.9\%)  \\

\hline
\end{tabular}
\subcaption{Number (percentage) of users per group}
\label{tab:gt_stats_user}

\end{subtable}%

\begin{subtable}{\columnwidth}
\centering
\begin{tabular}{cccc}
\hline

& Liberal    & Conservative \\
\hline
Humans      &957,726 (37.0\%)   & 476,231 (18.4\%)     \\
Bots      & 288,659 (11.1\%)    & 364,727 (14.1\%)  \\

\hline
\end{tabular}
\subcaption{Number (percentage) of tweets per group}
\label{tab:gt_stats_tweet}
\end{subtable}%
\end{table}



\begin{figure*}[!htb]
    \centering
    \resizebox{0.75\textwidth}{!}{%
    \begin{tikzpicture}
        \node[anchor=south west,inner sep=0] (image) at (0,0) {\includegraphics[width=\textwidth]{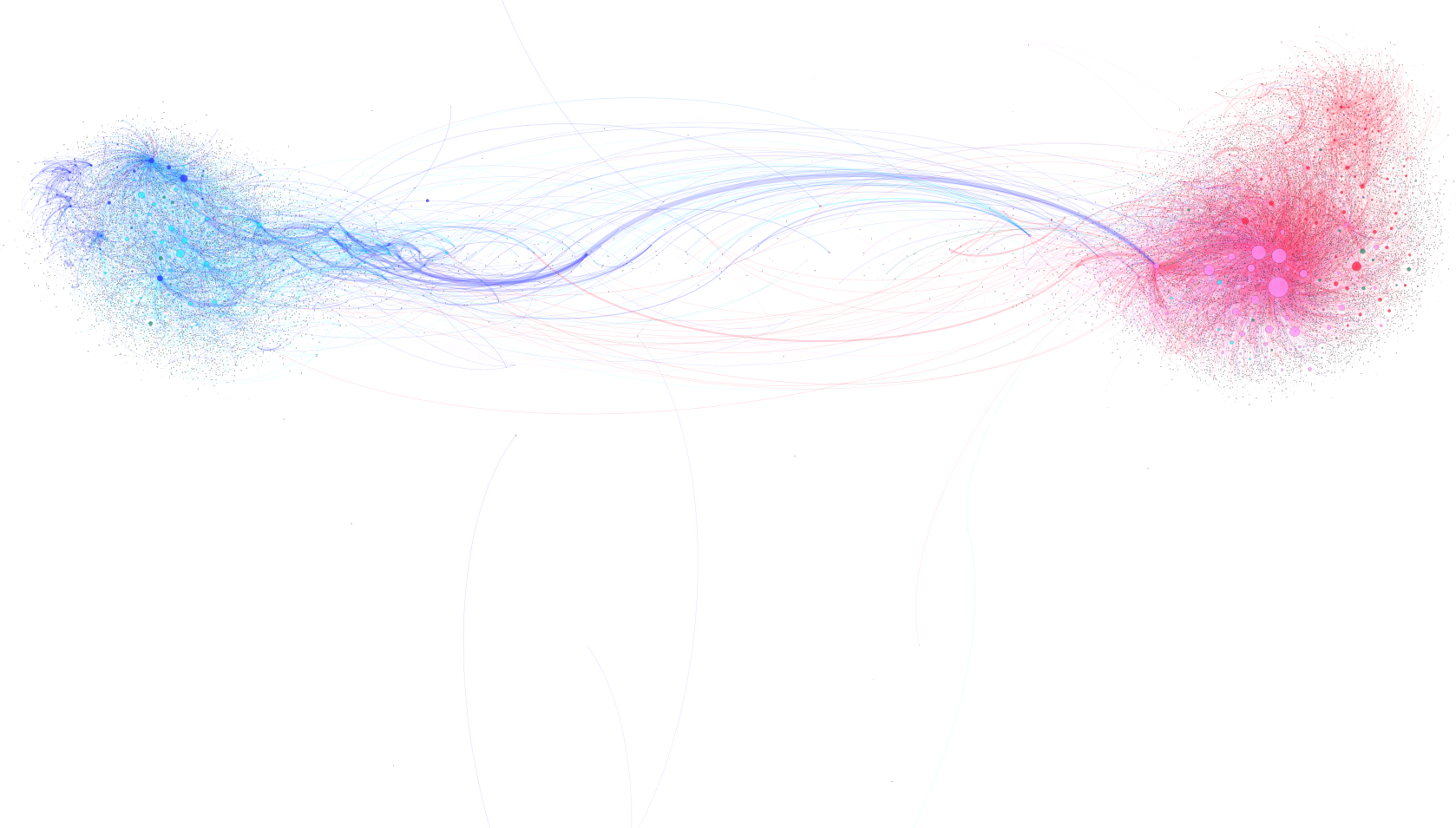}};
        \begin{scope}[x={(image.south east)},y={(image.north west)}]

            \node[anchor=south west,inner sep=0] (image) at (0.0,0.05) {\includegraphics[width=\textwidth]{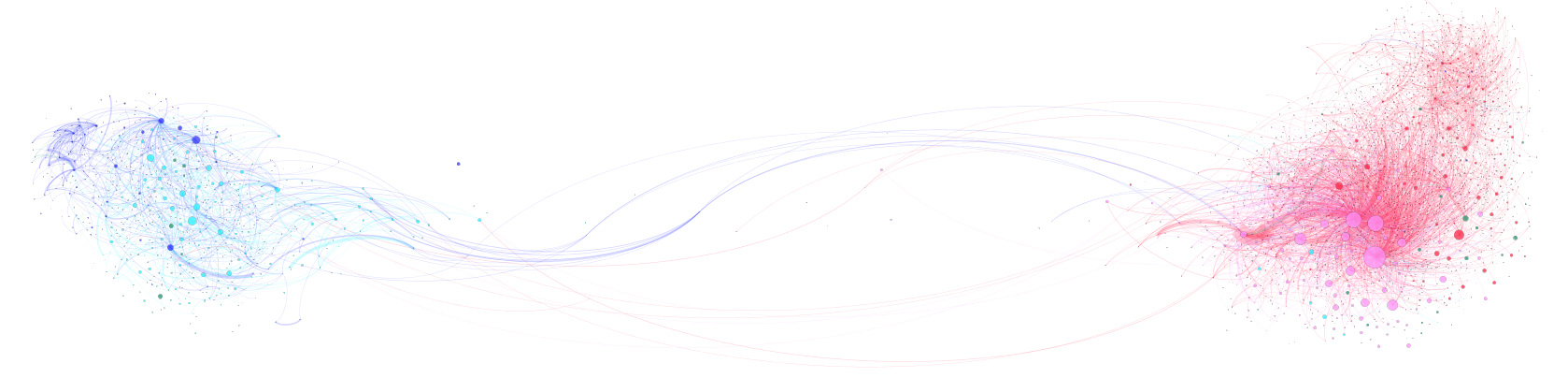}};
        \end{scope}
\node[draw=none] at (0,1) {\large \textit{(b) 25-core decomposition}};
\node[draw=none] at (0,5.5) {\large \textit{(a) 10-core decomposition}};

    \end{tikzpicture}
    }%
    \caption{Political discussion over \textit{(a)} the 10-core, and \textit{(b)} the 25-core decomposition of the retweet network. Each node represents a user, while links represent retweets. Links with weight (i.e., frequency of occurrence) less than 2 are hidden to minimize visual clutter. Blue  nodes represent liberal accounts, while red nodes indicate conservative users. Darker tones (blue and red) depict bots, while lighter tones (cyan and pink) relate to humans, and the few green nodes represent unclassified accounts. The link takes the same color of the source node (author of the retweet), whereas node size is proportional to the in-degree of the user.}
    \label{fig:bot_score_comparison}
\end{figure*}


\begin{table}[t!]
\centering\small
\caption{Top 20 hashtags generated by liberal and conservative bots.
Hashtags in bold are not present in the top 50 hashtags used by the corresponding human group.} 
\label{tab:bots_hash}
\begin{tabular}{cc}
\hline

 Liberal Bots & Conservative Bots  \\
\hline

\#MAGA & \#BrowardCounty \\ 
\#NovemberisComing & \#MAGA \\ 
\#TheResistance & \#StopTheSteal\\
\#GOTV & \#WalkAway \\
\#Florida & \#WednesdayWisdom \\
\textbf{\#ImpeachTrump} & \#PalmBeachCounty \\
\#Russia &  \#Florida\\  
\#VoteThemOut & \#QAnon\\ 
\#unhackthevote & \#KAG\\
\textbf{\#FlipTheHouse}   & \#IranRegime \\
\#RegisterToVote & \#Tehran \\
\#Resist & \#WWG1WGA\\
\textbf{\#ImpeachKavanaugh} & \textbf{\#Louisiana}\\
\#GOP & \#BayCounty\\
\#MeToo & \#AmericaFirst\\
\#AMJoy & \#DemocratsAreDangerous \\
\#txlege & \#StopTheCaravan \\
\textbf{\#FlipTheSenate} & \textbf{\#Blexit} \\
\textbf{\#CultureOfCorruption} & \textbf{\#VoteDemsOut} \\
\textbf{\#TrumpTrain} & \#VoterFraud \\

\hline
\end{tabular}
\end{table}

The combination of the outcome from the bot detection algorithm and the political ideology inference allowed us to identify four groups of users, namely Liberal Humans, Conservative Humans, Liberal Bots, and Conservative Bots.
In Table \ref{tab:gt_stats_user}, we show the percentage of users per group. Note that percentages do not sum up to 100 as either the political ideology inference was not able to classify every user, or Botometer did not return a score, as we previously mentioned.
In particular, we were able to assign a political leaning to 63\% of bots and 67\% of humans.
We find that the liberal user population is almost three times larger than the conservative counterpart.
This discrepancy is also present, but less evident, for the bot accounts, which exhibit an unbalance in favor of liberal bots.
Further, we investigate the suspended accounts to inspect the consistency of this result. The inference algorithm attributed a political ideology to 63\% of these accounts, which in turn show once again the liberal advantage over the conservative faction (45\% vs. 18\%).

Figure \ref{fig:bot_score_comparison} shows two \textit{k}-core decomposition graphs of the retweet network. 
In a \textit{k}-core, each node is connected with at least \textit{k} other nodes. 
Figures \ref{fig:bot_score_comparison}a and \ref{fig:bot_score_comparison}b capture the 10-core and 25-core decomposition, respectively.
Here, nodes represent Twitter users and link represent retweets among them. 
We indicate as \textit{source} the user that retweeted the tweet of a \textit{target} user.
Colors represent the political ideology, with darker colors (red and blue) being bots and lighter colors (cyan and pink) being human users; size represents the in-degree.
The graph is visualized using a force-directed layout \cite{jacomy2014forceatlas2}, where nodes repulse each other, while edges attract their nodes.
In our setting, this means that users are spatially distributed according to the amount of retweets between each other.
The result is a network naturally split into two communities, where each side is almost entirely populated by users with the same political ideology. %
This polarization is also reflected by bots, which are embedded, with humans, in each political side. Two facts are worth noting: \textit{(i)} as \textit{k} increases, the left k-core appears to disrupt, while the right k-core remains well connected; and, \textit{(ii)} as \textit{k} increases, bots appear to outnumber humans, suggesting that bots may populate areas of the retweet network that are more central and better connected. 

Next, we examine the topics discussed by social bots and compare them with the human counterparts.
Table \ref{tab:bots_hash} shows the top 20 hashtags utilized by liberal and conservative bots.
We highlight (in bold) the hashtags that are not present in the top 50 hashtags used by the corresponding human group to point out the similarities and differences among the groups.
In this table, we do not take into account general hashtags (such as \#elections, \#midterms, \#democrats, \#liberals, \#VoteRed(or Blue)ToSaveAmerica, and \#Trump) as $(i)$ the overlap between bot and human hashtags is noticeable when these terms are considered, and $(ii)$ we aim to narrow the analysis to specific topics and inflammatory content, inspired by \cite{stella2018bots}. 
Moreover, we used an enlarged subset of hashtags for the human groups to further strengthen the differences and, at the same time, to better understand the objective of social bots.
Although bots and humans share the majority of hashtags, two main differences can be noticed.
First, conservative bots abide by the  corresponding human counterpart more than the liberal bots.
Second, liberal bots focus on more inflammatory and provocative content (e.g., \#ImpeachTrump, \#ImpeachKavanaugh, \#FlipTheSenate) w.r.t.  conservative bots.


\subsection{RQ2: Bot Activity and Strategies}
In this Section, we investigate social bot activity based on their political leaning. We explore their strategies in interacting with humans and the degree of embeddedness in the social network.




Table \ref{tab:gt_stats_tweet} depicts the number (and percentage) of tweets generated by each group. 
Despite the group composed of conservative bots is the smallest in terms of number of accounts, it produced more tweets than liberal bots and closely approaches the number of tweets generated by the human counterpart. The resulting \textit{tweet per user ratio} shows that conservative bots produce 7.4 tweets per account, which is more than twice the ratio related to the liberal bots (3.5), almost the double of the human counterpart (3.9), and nearly three times the ratio of liberal humans (2.5).

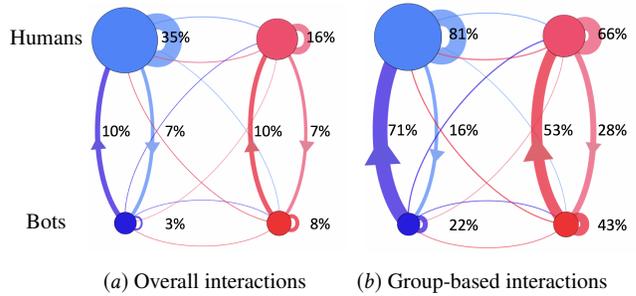
\begin{figure}[!t]
    \centering
    \resizebox{\columnwidth}{!}{%
    \begin{tikzpicture}
        \node[anchor=south west,inner sep=0] (image) at (0,0) {\includegraphics[width=.47\columnwidth]{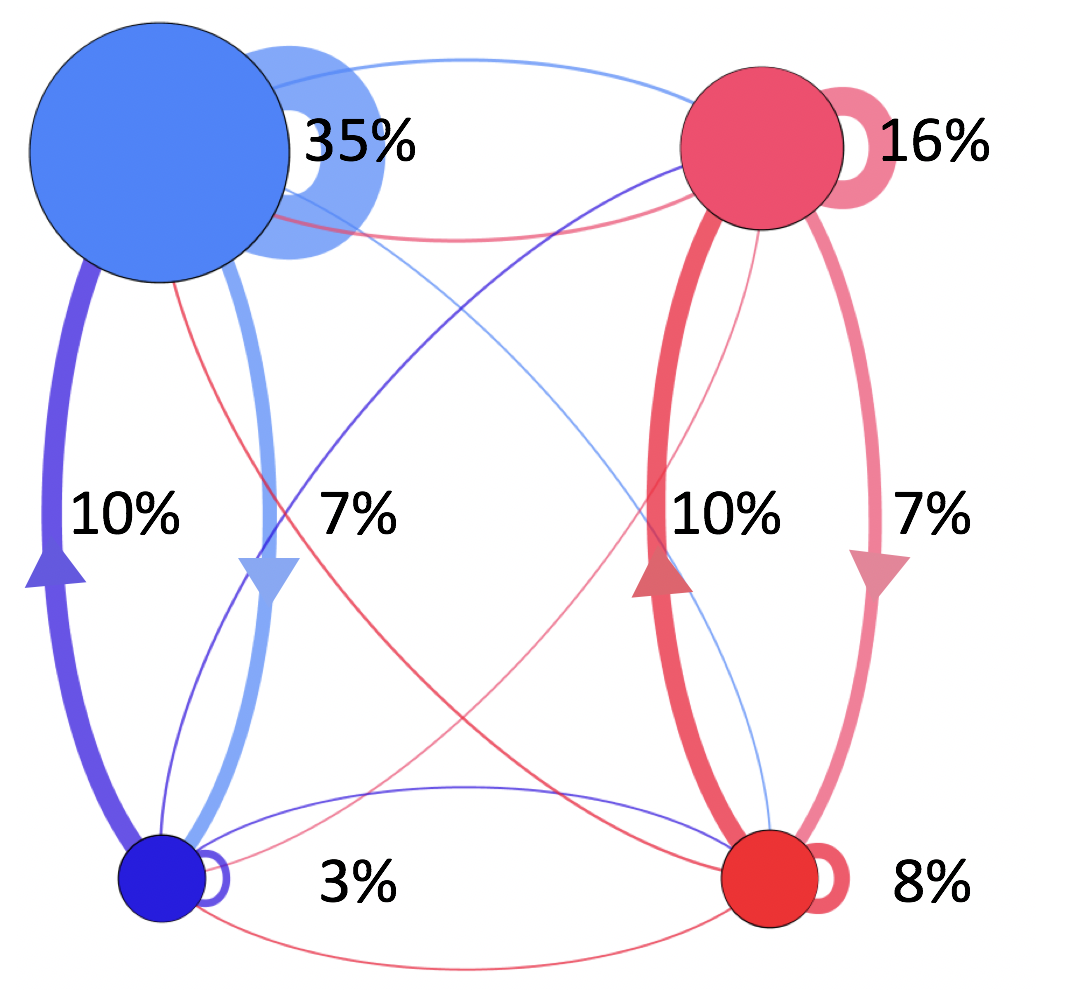}};
        \begin{scope}[x={(image.south east)},y={(image.north west)}]

            \node[anchor=south west,inner sep=0] (image) at (1,-.02) {\includegraphics[width=.47\columnwidth]{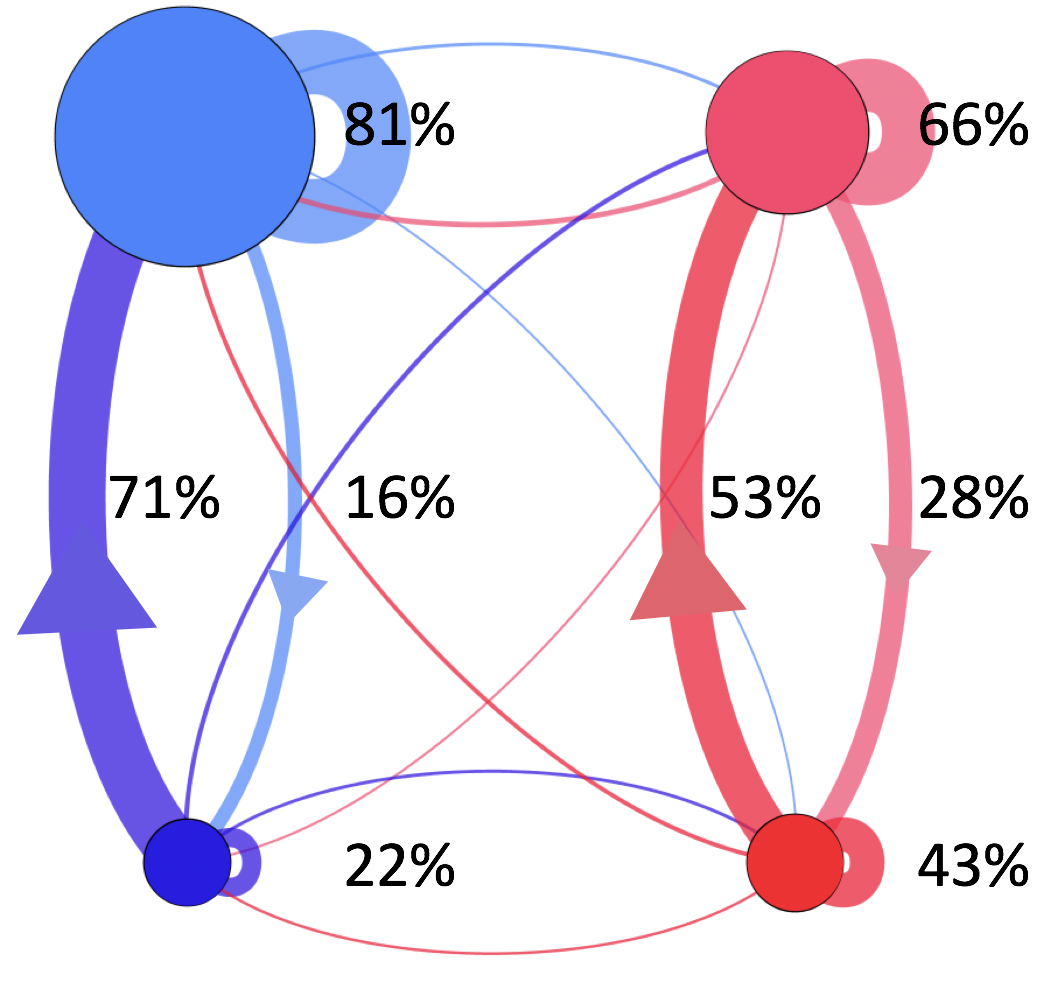}};
        \end{scope}
\node[draw=none] at (-.55,0.485) {Bots};
\node[draw=none] at (-.55,3.15) {Humans};
\node[draw=none] at (1.75,-0.4) {$(a)$ Overall interactions};
\node[draw=none] at (5.775,-0.4) {$(b)$ Group-based interactions};

    \end{tikzpicture}
    }%
    \caption{Interactions according to political ideology}
    \label{fig:rt_pid_overall}
\end{figure}

To investigate the interplay between bots and humans, we consider the previously described retweet network. 
Figure \ref{fig:rt_pid_overall} shows the interaction among the four groups.
We maintain the same color mapping described before, with darker color (on the bottom) representing bots and lighter color (on top) indicating humans. Node size is proportional to the percentage of accounts in each group, while edge size is proportional to the percentage of interactions between each group. 
In Figure \ref{fig:rt_pid_overall}a, this percentage is computed considering all the interactions in the retweet network, while in Figure \ref{fig:rt_pid_overall}b we consider each group separately, therefore, the edge size gives a measure of the group propensity to interact with the other groups.
Consistently with Figure \ref{fig:bot_score_comparison}, we observe that there is a limited amount of interaction between the two political sides. The majority of interactions are either intra-group or between groups of the same political leaning.
From Figure \ref{fig:rt_pid_overall}b, we can observe that the two bot factions adopted different strategies.
Conservative bots balanced their interactions by retweeting group members 43\% of the time, and the human counterpart 52\% of the time.
On the other hand, liberal bots mainly retweeted liberal humans (71\% of the time) and limited the intra-group interactions to the 22\% of their retweet activity.
Interestingly, conservative humans interacted with the conservative bots (28\% of the time) much more than the liberal counterpart (16\%) with the liberal bots. 
To better understand these results and to measure the effectiveness of both the strategies, in the next Section we evaluate the four metrics introduced earlier in this paper.

\begin{table}[t!]
\caption{Average network centrality measures}
\label{tab:in_n_out}
\centering\small
\begin{subtable}{.48\columnwidth}
\centering
\begin{tabular}{ccc}
\hline

& Liberal    & Conservative \\
\hline
Humans      & 2.66 $\cdot 10^{-6}$    & 4.14 $\cdot 10^{-6}$      \\
Bots      & 3.70 $\cdot 10^{-6}$     & 7.81 $\cdot 10^{-6}$   \\

\hline
\end{tabular}
\subcaption{Out-degree centrality}
\label{tab:out_deg}
\end{subtable}%

\begin{subtable}{.48\columnwidth}
\centering
\begin{tabular}{ccc}
\hline

& Liberal    & Conservative \\
\hline
Humans      & 2.52 $\cdot 10^{-6}$    & 4.24 $\cdot 10^{-6}$      \\
Bots      & 2.53 $\cdot 10^{-6}$     & 6.22 $\cdot 10^{-6}$   \\

\hline
\end{tabular}
\subcaption{In-degree centrality}
\label{tab:in_deg}
\end{subtable}%
\vspace{-3 mm}
\end{table}

\begin{figure}[t] 
\centering
  \includegraphics[width=\columnwidth]{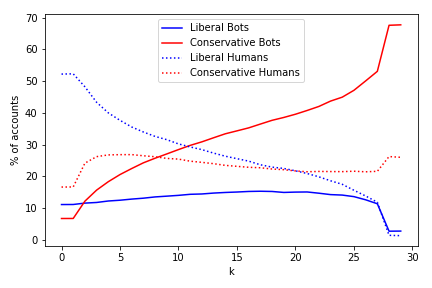}
  \caption{k-core decomposition, liberal vs. conservative users}
  \label{fig:core}
\vspace{-.35cm}
\end{figure}

Finally, we examine the degree of embeddedness of both humans and bots within the retweet network.
For this purpose, we first compute different network centrality measures, and then we adopt the k-core decomposition technique to identify the most central nodes in the graph.
In Table \ref{tab:in_n_out}, we show the average out- and in-degree centrality for each group of users. 
Out-degree centrality measures the quantity of outgoing links, while in-degree centrality considers the number of of incoming links. Both of these measures are normalized by the maximum possible degree of the graph.
Overall, conservative groups have higher centrality measures than the liberal ones.
We can notice that conservative bots achieve the highest values both for the out- and in-degree centrality.
To further investigate bots embeddedness in the social network, we use the k-core decomposition. The objective of this technique is to determine the set of nodes deeply embedded in a graph. The k-core is a subgraph of the original graph in which every node has a degree equal to or greater than a given value $k$.
We extracted the k-cores from the retweet network by varying $k$ in the range between 0 and 30.
Figure \ref{fig:core} depicts the percentage of liberal and conservative users as a function of $k$.
We can notice that, as $k$ grows, the fraction of conservative bots increases, while the percentage of liberal bots remains almost stationary.
On the human side, the liberal fraction drops with $k$, whereas the conservative percentage remains approximately steady.  
Overall, conservative bots sit in a more central position in the social network and are more deeply connected if compared to the liberal counterpart.

\subsection{RQ3: Bot Effectiveness}
\begin{table}[t]
\centering\small
\caption{Bot effectiveness
\label{tab:eff}} 
\begin{tabular}{cccc}
\hline

Metric & Liberal Bots & Conservative Bots  \\
\hline
$RTP$     & 14.1\% & 25.6\% \\
$RR$      &4.5\% & 15.5\%\\
$H2BR$    &12.3\% & 23.2\%\\
$TSR$     &35.3\% & 35.0\%\\

\hline
\end{tabular}
\end{table}

In this Section, we aim to estimate the effectiveness of bot strategies and measure to what extent humans rely upon, and interact with the content generated by social bots.
We examine the effect of bot activities by means of the four metrics described in Section \textit{Bot Activity Effectiveness}.
We evaluate each political side separately, thus, we compare the interaction between bots and humans with the same leaning. In Table \ref{tab:eff}, we depict the results for each group of bots.
Diverse aspects are worthy of consideration.
We can observe that conservative bots are significantly more effective than the liberal counterpart.
Although the $TSR$s of the \textit{red} and \textit{blue} bots are comparable, the gap between the two groups, with respect to the other metrics, is significant.
To carefully interpret this result, it should also be noticed that $(i)$ the $TSR$ is inversely proportional to the number of tweets generated by bots, and $(ii)$ conservative bots tweeted more than the liberal counterpart, as depicted in Table \ref{tab:gt_stats_tweet}.
Overall, conservative bots received a larger degree of interaction with (and likely trust from) human users.
In fact, conservative humans interacted with the bot counterpart almost twice with retweets  ($RTP$), and more than three times with replies ($RR$) if compared to the liberal group.
Finally, the $H2BR$ highlights a remarkable amount of human activities that involve social bots: almost one in four actions performed by conservative humans goes towards red bots.

%% file: src/conclusion.tex
\section{Conclusions \& Future Work}

In this work, we conducted an investigation to analyze social bots activity during the 2018 US Midterm election. 
We showed that social bots are embedded in each political wing and behave accordingly.
We observed different strategies between conservative and liberal bots. Specifically, conservative bots stand in a more central position in the social network and abide by the topic discussed by the human counterpart more than the liberal bots, which in turn exhibit an inflammatory attitude.
Further, conservative bots balanced their interaction with humans and bots of the red wing, whereas liberal bots focused mainly on the interplay with the human counterpart.

Finally, we inspected the effectiveness of these strategies and recognized the strategy of the conservative bots as the most effective.
However, these results open the door to further interpretation and discussion.
Are conservative bots more effective because of their strategy or because of the human ineptitude to distinguish their nature?
This, and related analysis, will be expanded  in  future work.

%% file: main.bbl

\begin{thebibliography}{31}


\ifx \showCODEN    \undefined \def \showCODEN     #1{\unskip}     \fi
\ifx \showDOI      \undefined \def \showDOI       #1{#1}\fi
\ifx \showISBNx    \undefined \def \showISBNx     #1{\unskip}     \fi
\ifx \showISBNxiii \undefined \def \showISBNxiii  #1{\unskip}     \fi
\ifx \showISSN     \undefined \def \showISSN      #1{\unskip}     \fi
\ifx \showLCCN     \undefined \def \showLCCN      #1{\unskip}     \fi
\ifx \shownote     \undefined \def \shownote      #1{#1}          \fi
\ifx \showarticletitle \undefined \def \showarticletitle #1{#1}   \fi
\ifx \showURL      \undefined \def \showURL       {\relax}        \fi
\providecommand\bibfield[2]{#2}
\providecommand\bibinfo[2]{#2}
\providecommand\natexlab[1]{#1}
\providecommand\showeprint[2][]{arXiv:#2}

\bibitem[\protect\citeauthoryear{Badawy, Ferrara, and Lerman}{Badawy
  et~al\mbox{.}}{2018a}]%
        {Badawy2018}
\bibfield{author}{\bibinfo{person}{Adam Badawy}, \bibinfo{person}{Emilio
  Ferrara}, {and} \bibinfo{person}{Kristina Lerman}.}
  \bibinfo{year}{2018}\natexlab{a}.
\newblock \showarticletitle{Analyzing the Digital Traces of Political
  Manipulation: The 2016 Russian Interference Twitter Campaign}. In
  \bibinfo{booktitle}{\emph{Int. Conference on Advances in Social Networks
  Analysis and Mining}}. \bibinfo{pages}{258--265}.
\newblock


\bibitem[\protect\citeauthoryear{Badawy, Lerman, and Ferrara}{Badawy
  et~al\mbox{.}}{2018b}]%
        {badawy2018falls}
\bibfield{author}{\bibinfo{person}{Adam Badawy}, \bibinfo{person}{Kristina
  Lerman}, {and} \bibinfo{person}{Emilio Ferrara}.}
  \bibinfo{year}{2018}\natexlab{b}.
\newblock \showarticletitle{Who Falls for Online Political Manipulation?}
\newblock \bibinfo{journal}{\emph{arXiv preprint arXiv:1808.03281}}
  (\bibinfo{year}{2018}).
\newblock


\bibitem[\protect\citeauthoryear{Bessi and Ferrara}{Bessi and Ferrara}{2016}]%
        {bessi2016social}
\bibfield{author}{\bibinfo{person}{Alessandro Bessi} {and}
  \bibinfo{person}{Emilio Ferrara}.} \bibinfo{year}{2016}\natexlab{}.
\newblock \showarticletitle{Social bots distort the 2016 US Presidential
  election online discussion}.
\newblock \bibinfo{journal}{\emph{First Monday}} \bibinfo{volume}{21},
  \bibinfo{number}{11} (\bibinfo{year}{2016}).
\newblock


\bibitem[\protect\citeauthoryear{Boichak, Jackson, Hemsley, and
  Tanupabrungsun}{Boichak et~al\mbox{.}}{2018}]%
        {boichak2018automated}
\bibfield{author}{\bibinfo{person}{Olga Boichak}, \bibinfo{person}{Sam
  Jackson}, \bibinfo{person}{Jeff Hemsley}, {and} \bibinfo{person}{Sikana
  Tanupabrungsun}.} \bibinfo{year}{2018}\natexlab{}.
\newblock \showarticletitle{Automated Diffusion? Bots and Their Influence
  During the 2016 US Presidential Election}. In
  \bibinfo{booktitle}{\emph{International Conference on Information}}.
  Springer, \bibinfo{pages}{17--26}.
\newblock


\bibitem[\protect\citeauthoryear{Bovet and Makse}{Bovet and Makse}{2019}]%
        {bovet2019influence}
\bibfield{author}{\bibinfo{person}{Alexandre Bovet} {and}
  \bibinfo{person}{Hern{\'a}n~A Makse}.} \bibinfo{year}{2019}\natexlab{}.
\newblock \showarticletitle{Influence of fake news in Twitter during the 2016
  US presidential election}.
\newblock \bibinfo{journal}{\emph{Nature communications}} \bibinfo{volume}{10},
  \bibinfo{number}{1} (\bibinfo{year}{2019}), \bibinfo{pages}{7}.
\newblock


\bibitem[\protect\citeauthoryear{Breiman}{Breiman}{2001}]%
        {breiman2001random}
\bibfield{author}{\bibinfo{person}{Leo Breiman}.}
  \bibinfo{year}{2001}\natexlab{}.
\newblock \showarticletitle{Random forests}.
\newblock \bibinfo{journal}{\emph{Machine learning}} \bibinfo{volume}{45},
  \bibinfo{number}{1} (\bibinfo{year}{2001}), \bibinfo{pages}{5--32}.
\newblock


\bibitem[\protect\citeauthoryear{Chavoshi, Hamooni, and Mueen}{Chavoshi
  et~al\mbox{.}}{2016}]%
        {chavoshi2016debot}
\bibfield{author}{\bibinfo{person}{Nikan Chavoshi}, \bibinfo{person}{Hossein
  Hamooni}, {and} \bibinfo{person}{Abdullah Mueen}.}
  \bibinfo{year}{2016}\natexlab{}.
\newblock \showarticletitle{DeBot: Twitter Bot Detection via Warped
  Correlation.}. In \bibinfo{booktitle}{\emph{ICDM}}.
  \bibinfo{pages}{817--822}.
\newblock


\bibitem[\protect\citeauthoryear{Chen and Subramanian}{Chen and
  Subramanian}{2018}]%
        {chen2018unsupervised}
\bibfield{author}{\bibinfo{person}{Zhouhan Chen} {and} \bibinfo{person}{Devika
  Subramanian}.} \bibinfo{year}{2018}\natexlab{}.
\newblock \showarticletitle{An Unsupervised Approach to Detect Spam Campaigns
  that Use Botnets on Twitter}.
\newblock \bibinfo{journal}{\emph{arXiv preprint arXiv:1804.05232}}
  (\bibinfo{year}{2018}).
\newblock


\bibitem[\protect\citeauthoryear{Davis, Varol, Ferrara, Flammini, and
  Menczer}{Davis et~al\mbox{.}}{2016}]%
        {davis2016botornot}
\bibfield{author}{\bibinfo{person}{Clayton~Allen Davis}, \bibinfo{person}{Onur
  Varol}, \bibinfo{person}{Emilio Ferrara}, \bibinfo{person}{Alessandro
  Flammini}, {and} \bibinfo{person}{Filippo Menczer}.}
  \bibinfo{year}{2016}\natexlab{}.
\newblock \showarticletitle{Botornot: A system to evaluate social bots}. In
  \bibinfo{booktitle}{\emph{Proceedings of the 25th International Conference
  Companion on World Wide Web}}.
\newblock


\bibitem[\protect\citeauthoryear{Deb, Luceri, Badawy, and Ferrara}{Deb
  et~al\mbox{.}}{2019}]%
        {deb2019bots}
\bibfield{author}{\bibinfo{person}{Ashok Deb}, \bibinfo{person}{Luca Luceri},
  \bibinfo{person}{Adam Badawy}, {and} \bibinfo{person}{Emilio Ferrara}.}
  \bibinfo{year}{2019}\natexlab{}.
\newblock \showarticletitle{Perils and Challenges of Social Media and Election
  Manipulation Analysis: The 2018 US Midterms}.
\newblock \bibinfo{journal}{\emph{arXiv preprint arXiv:1902.00043}}
  (\bibinfo{year}{2019}).
\newblock


\bibitem[\protect\citeauthoryear{Ferrara}{Ferrara}{2015}]%
        {ferrara2015manipulation}
\bibfield{author}{\bibinfo{person}{Emilio Ferrara}.}
  \bibinfo{year}{2015}\natexlab{}.
\newblock \showarticletitle{Manipulation and abuse on social media}.
\newblock \bibinfo{journal}{\emph{ACM SIGWEB Newsletter}}
  \bibinfo{number}{Spring} (\bibinfo{year}{2015}), \bibinfo{pages}{4}.
\newblock


\bibitem[\protect\citeauthoryear{Ferrara}{Ferrara}{2017}]%
        {ferrara2017disinformation}
\bibfield{author}{\bibinfo{person}{Emilio Ferrara}.}
  \bibinfo{year}{2017}\natexlab{}.
\newblock \showarticletitle{Disinformation and Social Bot Operations in the Run
  Up to the 2017 French Presidential Election}.
\newblock \bibinfo{journal}{\emph{First Monday}} \bibinfo{volume}{22},
  \bibinfo{number}{8} (\bibinfo{year}{2017}).
\newblock


\bibitem[\protect\citeauthoryear{Ferrara, Varol, Davis, Menczer, and
  Flammini}{Ferrara et~al\mbox{.}}{2016}]%
        {ferrara2016rise}
\bibfield{author}{\bibinfo{person}{Emilio Ferrara}, \bibinfo{person}{Onur
  Varol}, \bibinfo{person}{Clayton Davis}, \bibinfo{person}{Filippo Menczer},
  {and} \bibinfo{person}{Alessandro Flammini}.}
  \bibinfo{year}{2016}\natexlab{}.
\newblock \showarticletitle{The rise of social bots}.
\newblock \bibinfo{journal}{\emph{Commun. ACM}} \bibinfo{volume}{59},
  \bibinfo{number}{7} (\bibinfo{year}{2016}), \bibinfo{pages}{96--104}.
\newblock


\bibitem[\protect\citeauthoryear{Grinberg, Joseph, Friedland, Swire-Thompson,
  and Lazer}{Grinberg et~al\mbox{.}}{2019}]%
        {Grinberg2019}
\bibfield{author}{\bibinfo{person}{Nir Grinberg}, \bibinfo{person}{Kenneth
  Joseph}, \bibinfo{person}{Lisa Friedland}, \bibinfo{person}{Briony
  Swire-Thompson}, {and} \bibinfo{person}{David Lazer}.}
  \bibinfo{year}{2019}\natexlab{}.
\newblock \showarticletitle{{Fake news on Twitter during the 2016 U.S.
  presidential election}}.
\newblock \bibinfo{journal}{\emph{Science}} \bibinfo{volume}{363},
  \bibinfo{number}{6425} (\bibinfo{year}{2019}), \bibinfo{pages}{374--378}.
\newblock


\bibitem[\protect\citeauthoryear{Guess, Nagler, and Tucker}{Guess
  et~al\mbox{.}}{2019}]%
        {guess2019less}
\bibfield{author}{\bibinfo{person}{Andrew Guess}, \bibinfo{person}{Jonathan
  Nagler}, {and} \bibinfo{person}{Joshua Tucker}.}
  \bibinfo{year}{2019}\natexlab{}.
\newblock \showarticletitle{Less than you think: Prevalence and predictors of
  fake news dissemination on Facebook}.
\newblock \bibinfo{journal}{\emph{Science Advances}} \bibinfo{volume}{5},
  \bibinfo{number}{1} (\bibinfo{year}{2019}), \bibinfo{pages}{eaau4586}.
\newblock


\bibitem[\protect\citeauthoryear{Howard, Bolsover, Kollanyi, Bradshaw, and
  Neudert}{Howard et~al\mbox{.}}{2017}]%
        {howard2017junk}
\bibfield{author}{\bibinfo{person}{Philip~N Howard}, \bibinfo{person}{Gillian
  Bolsover}, \bibinfo{person}{Bence Kollanyi}, \bibinfo{person}{Samantha
  Bradshaw}, {and} \bibinfo{person}{Lisa-Maria Neudert}.}
  \bibinfo{year}{2017}\natexlab{}.
\newblock \showarticletitle{Junk news and bots during the US election: What
  were Michigan voters sharing over Twitter}.
\newblock \bibinfo{journal}{\emph{CompProp, OII, Data Memo}}
  (\bibinfo{year}{2017}).
\newblock


\bibitem[\protect\citeauthoryear{Im, Chandrasekharan, Sargent, Lighthammer,
  Denby, Bhargava, Hemphill, Jurgens, and Gilbert}{Im et~al\mbox{.}}{2019}]%
        {im2019still}
\bibfield{author}{\bibinfo{person}{Jane Im}, \bibinfo{person}{Eshwar
  Chandrasekharan}, \bibinfo{person}{Jackson Sargent}, \bibinfo{person}{Paige
  Lighthammer}, \bibinfo{person}{Taylor Denby}, \bibinfo{person}{Ankit
  Bhargava}, \bibinfo{person}{Libby Hemphill}, \bibinfo{person}{David Jurgens},
  {and} \bibinfo{person}{Eric Gilbert}.} \bibinfo{year}{2019}\natexlab{}.
\newblock \showarticletitle{Still out there: Modeling and Identifying Russian
  Troll Accounts on Twitter}.
\newblock \bibinfo{journal}{\emph{arXiv preprint arXiv:1901.11162}}
  (\bibinfo{year}{2019}).
\newblock


\bibitem[\protect\citeauthoryear{Jacomy, Venturini, Heymann, and
  Bastian}{Jacomy et~al\mbox{.}}{2014}]%
        {jacomy2014forceatlas2}
\bibfield{author}{\bibinfo{person}{Mathieu Jacomy}, \bibinfo{person}{Tommaso
  Venturini}, \bibinfo{person}{Sebastien Heymann}, {and}
  \bibinfo{person}{Mathieu Bastian}.} \bibinfo{year}{2014}\natexlab{}.
\newblock \showarticletitle{ForceAtlas2, a continuous graph layout algorithm
  for handy network visualization designed for the Gephi software}.
\newblock \bibinfo{journal}{\emph{PloS one}} \bibinfo{volume}{9},
  \bibinfo{number}{6} (\bibinfo{year}{2014}), \bibinfo{pages}{e98679}.
\newblock


\bibitem[\protect\citeauthoryear{Kudugunta and Ferrara}{Kudugunta and
  Ferrara}{2018}]%
        {kudugunta2018deep}
\bibfield{author}{\bibinfo{person}{Sneha Kudugunta} {and}
  \bibinfo{person}{Emilio Ferrara}.} \bibinfo{year}{2018}\natexlab{}.
\newblock \showarticletitle{Deep Neural Networks for Bot Detection}.
\newblock \bibinfo{journal}{\emph{Information Sciences}} \bibinfo{volume}{467},
  \bibinfo{number}{October} (\bibinfo{year}{2018}), \bibinfo{pages}{312--322}.
\newblock


\bibitem[\protect\citeauthoryear{M{\o}nsted, Sapie{\.z}y{\'n}ski, Ferrara, and
  Lehmann}{M{\o}nsted et~al\mbox{.}}{2017}]%
        {monsted2017evidence}
\bibfield{author}{\bibinfo{person}{Bjarke M{\o}nsted}, \bibinfo{person}{Piotr
  Sapie{\.z}y{\'n}ski}, \bibinfo{person}{Emilio Ferrara}, {and}
  \bibinfo{person}{Sune Lehmann}.} \bibinfo{year}{2017}\natexlab{}.
\newblock \showarticletitle{Evidence of Complex Contagion of Information in
  Social Media: An Experiment Using Twitter Bots}.
\newblock \bibinfo{journal}{\emph{Plos One}} \bibinfo{volume}{12},
  \bibinfo{number}{9} (\bibinfo{year}{2017}), \bibinfo{pages}{e0184148}.
\newblock


\bibitem[\protect\citeauthoryear{Persily}{Persily}{2017}]%
        {persily20172016}
\bibfield{author}{\bibinfo{person}{Nathaniel Persily}.}
  \bibinfo{year}{2017}\natexlab{}.
\newblock \showarticletitle{The 2016 US Election: Can democracy survive the
  internet?}
\newblock \bibinfo{journal}{\emph{Journal of democracy}} \bibinfo{volume}{28},
  \bibinfo{number}{2} (\bibinfo{year}{2017}), \bibinfo{pages}{63--76}.
\newblock


\bibitem[\protect\citeauthoryear{Pozzana and Ferrara}{Pozzana and
  Ferrara}{2018}]%
        {pozzana2018measuring}
\bibfield{author}{\bibinfo{person}{Iacopo Pozzana} {and}
  \bibinfo{person}{Emilio Ferrara}.} \bibinfo{year}{2018}\natexlab{}.
\newblock \showarticletitle{Measuring bot and human behavioral dynamics}.
\newblock \bibinfo{journal}{\emph{arXiv preprint arXiv:1802.04286}}
  (\bibinfo{year}{2018}).
\newblock


\bibitem[\protect\citeauthoryear{Scheufele and Krause}{Scheufele and
  Krause}{2019}]%
        {scheufele2019science}
\bibfield{author}{\bibinfo{person}{Dietram~A Scheufele} {and}
  \bibinfo{person}{Nicole~M Krause}.} \bibinfo{year}{2019}\natexlab{}.
\newblock \showarticletitle{Science audiences, misinformation, and fake news}.
\newblock \bibinfo{journal}{\emph{PNAS}} (\bibinfo{year}{2019}),
  \bibinfo{pages}{201805871}.
\newblock


\bibitem[\protect\citeauthoryear{Shao, Ciampaglia, Varol, Yang, Flammini, and
  Menczer}{Shao et~al\mbox{.}}{2018}]%
        {shao2018spread}
\bibfield{author}{\bibinfo{person}{Chengcheng Shao},
  \bibinfo{person}{Giovanni~Luca Ciampaglia}, \bibinfo{person}{Onur Varol},
  \bibinfo{person}{Kai-Cheng Yang}, \bibinfo{person}{Alessandro Flammini},
  {and} \bibinfo{person}{Filippo Menczer}.} \bibinfo{year}{2018}\natexlab{}.
\newblock \showarticletitle{The spread of low-credibility content by social
  bots}.
\newblock \bibinfo{journal}{\emph{Nature communications}} \bibinfo{volume}{9},
  \bibinfo{number}{1} (\bibinfo{year}{2018}), \bibinfo{pages}{4787}.
\newblock


\bibitem[\protect\citeauthoryear{Shu, Sliva, Wang, Tang, and Liu}{Shu
  et~al\mbox{.}}{2017}]%
        {shu2017fake}
\bibfield{author}{\bibinfo{person}{Kai Shu}, \bibinfo{person}{Amy Sliva},
  \bibinfo{person}{Suhang Wang}, \bibinfo{person}{Jiliang Tang}, {and}
  \bibinfo{person}{Huan Liu}.} \bibinfo{year}{2017}\natexlab{}.
\newblock \showarticletitle{Fake news detection on social media: A data mining
  perspective}.
\newblock \bibinfo{journal}{\emph{ACM SIGKDD Explorations Newsletter}}
  \bibinfo{volume}{19}, \bibinfo{number}{1} (\bibinfo{year}{2017}),
  \bibinfo{pages}{22--36}.
\newblock


\bibitem[\protect\citeauthoryear{Stella, Ferrara, and De~Domenico}{Stella
  et~al\mbox{.}}{2018}]%
        {stella2018bots}
\bibfield{author}{\bibinfo{person}{Massimo Stella}, \bibinfo{person}{Emilio
  Ferrara}, {and} \bibinfo{person}{Manlio De~Domenico}.}
  \bibinfo{year}{2018}\natexlab{}.
\newblock \showarticletitle{Bots increase exposure to negative and inflammatory
  content in online social systems}.
\newblock \bibinfo{journal}{\emph{Proceedings of the National Academy of
  Sciences}} \bibinfo{volume}{115}, \bibinfo{number}{49}
  (\bibinfo{year}{2018}), \bibinfo{pages}{12435--12440}.
\newblock


\bibitem[\protect\citeauthoryear{Subrahmanian, Azaria, Durst, Kagan, Galstyan,
  Lerman, Zhu, Ferrara, Flammini, Menczer, et~al\mbox{.}}{Subrahmanian
  et~al\mbox{.}}{2016}]%
        {subrahmanian2016darpa}
\bibfield{author}{\bibinfo{person}{VS Subrahmanian}, \bibinfo{person}{Amos
  Azaria}, \bibinfo{person}{Skylar Durst}, \bibinfo{person}{Vadim Kagan},
  \bibinfo{person}{Aram Galstyan}, \bibinfo{person}{Kristina Lerman},
  \bibinfo{person}{Linhong Zhu}, \bibinfo{person}{Emilio Ferrara},
  \bibinfo{person}{Alessandro Flammini}, \bibinfo{person}{Filippo Menczer},
  {et~al\mbox{.}}} \bibinfo{year}{2016}\natexlab{}.
\newblock \showarticletitle{The DARPA Twitter Bot Challenge}.
\newblock \bibinfo{journal}{\emph{Computer}} \bibinfo{volume}{49},
  \bibinfo{number}{6} (\bibinfo{year}{2016}).
\newblock


\bibitem[\protect\citeauthoryear{Varol, Ferrara, Davis, Menczer, and
  Flammini}{Varol et~al\mbox{.}}{2017}]%
        {varol2017online}
\bibfield{author}{\bibinfo{person}{Onur Varol}, \bibinfo{person}{Emilio
  Ferrara}, \bibinfo{person}{Clayton~A Davis}, \bibinfo{person}{Filippo
  Menczer}, {and} \bibinfo{person}{Alessandro Flammini}.}
  \bibinfo{year}{2017}\natexlab{}.
\newblock \showarticletitle{Online human-bot interactions: Detection,
  estimation, and characterization}. In \bibinfo{booktitle}{\emph{Int. AAAI
  Conference on Web and Social Media}}. \bibinfo{pages}{280--289}.
\newblock


\bibitem[\protect\citeauthoryear{Vosoughi, Roy, and Aral}{Vosoughi
  et~al\mbox{.}}{2018}]%
        {vosoughi2018spread}
\bibfield{author}{\bibinfo{person}{Soroush Vosoughi}, \bibinfo{person}{Deb
  Roy}, {and} \bibinfo{person}{Sinan Aral}.} \bibinfo{year}{2018}\natexlab{}.
\newblock \showarticletitle{The spread of true and false news online}.
\newblock \bibinfo{journal}{\emph{Science}} \bibinfo{volume}{359},
  \bibinfo{number}{6380} (\bibinfo{year}{2018}), \bibinfo{pages}{1146--1151}.
\newblock


\bibitem[\protect\citeauthoryear{Woolley and Guilbeault}{Woolley and
  Guilbeault}{2017}]%
        {woolley2017computational}
\bibfield{author}{\bibinfo{person}{Samuel~C Woolley} {and}
  \bibinfo{person}{Douglas~R Guilbeault}.} \bibinfo{year}{2017}\natexlab{}.
\newblock \showarticletitle{Computational propaganda in the United States of
  America: Manufacturing consensus online}.
\newblock \bibinfo{journal}{\emph{Computational Propaganda Research Project}}
  (\bibinfo{year}{2017}), \bibinfo{pages}{22}.
\newblock


\bibitem[\protect\citeauthoryear{Yang, Varol, Davis, Ferrara, Flammini, and
  Menczer}{Yang et~al\mbox{.}}{2019}]%
        {yang2019arming}
\bibfield{author}{\bibinfo{person}{Kai-Cheng Yang}, \bibinfo{person}{Onur
  Varol}, \bibinfo{person}{Clayton~A Davis}, \bibinfo{person}{Emilio Ferrara},
  \bibinfo{person}{Alessandro Flammini}, {and} \bibinfo{person}{Filippo
  Menczer}.} \bibinfo{year}{2019}\natexlab{}.
\newblock \showarticletitle{Arming the public with artificial intelligence to
  counter social bots}.
\newblock \bibinfo{journal}{\emph{Human Behavior and Emerging Technologies}}
  (\bibinfo{year}{2019}), \bibinfo{pages}{e115}.
\newblock


\end{thebibliography}
